\documentclass[preprint,12pt,authoryear]{elsarticle}

\usepackage{amsmath, amsthm, amssymb, amsfonts, pstricks, wasysym, stmaryrd, enumerate, epsfig}
\usepackage{rotating}
\usepackage{hyperref}
\hypersetup{colorlinks=true, urlcolor=blue}
\usepackage{graphicx}
\usepackage{longtable}

\usepackage{enumerate}
\usepackage{booktabs}
\usepackage{longtable}
\usepackage{color}
\usepackage[margin= 1in]{geometry}
\usepackage{amssymb}
\journal{arXiv}

\begin{document}

\begin{frontmatter}


%% use the corref command within \author for corresponding author footnotes

\title{The Arrival of News and Return Jumps in Stock Markets: A Nonparametric Approach}

\author[mymainaddress]{Juho Kanniainen\corref{mycorrespondingauthor}}
\cortext[mycorrespondingauthor]{Corresponding author. Contact information: juho.kanniainen@tuni.fi, +358 40 707 4532}
\ead{juho.kanniainen@tuni.fi}

\author[mymainaddress]{Ye Yue}

\address[mymainaddress]{Tampere University\\P.O. Box 541, FI-33101 Tampere, Finland\\ \ \\This version: \today}

%\textsc{}

\begin{abstract}
This paper introduces a non-parametric framework to statistically examine how news events, such as company or macroeconomic announcements, contribute to the pre- and post-event jump dynamics of stock prices under the intraday seasonality of the news and jumps. We demonstrate our framework, which has several advantages over the existing methods, by using data for i) the S\&P 500 index ETF, SPY, with macroeconomic announcements and ii) Nasdaq Nordic Large-Cap stocks with scheduled and non-scheduled company announcements. We provide strong evidence that non-scheduled company announcements and some macroeconomic announcements contribute jumps that follow the releases and also some evidence for pre-jumps that precede the scheduled arrivals of public information, which may indicate non-gradual information leakage. Especially interim reports of Nordic large-cap companies are found containing important information to yield jumps in stock prices. Additionally, our results show that releases of unexpected information are not reacted to uniformly across Nasdaq Nordic markets, even if they are jointly operated and are based on the same exchange rules.

\end{abstract}


\end{frontmatter}



%% INTRODUCTION
\section{Introduction} \label{Sec:Introduction}

Jumps in stock prices have an important role in returns dynamics with substantial implications for asset management and options pricing\footnote{See, for example, \citep[][]{bates1996jumps,duffie2000transform,cont2009constant,kaeck2012volatility,yang2016jumps}.}. A number of alternative jump detection methods have been proposed\footnote{\cite[][]{huang2005relative,barndorff2006econometrics,andersen2007no,jiang2008testing,Lee2008,corsi2010threshold,lee2010detecting,ait2011testing}.}. At the same time, according to recent empirical research, jumps in stock prices are likely to occur shortly after information arrivals \citep[see especially][]{Lee2008,Lee2012,bradley2014analysts}. 

Even if return jumps (hereafter just 'jumps') are almost always associated with news events, it does not mean that all news events -- a number of important news announcements arrive every day -- are always related to jumps. To examine how the arrivals of certain information (such as non-scheduled company announcements or macroeconomic announcements) contribute to the jumps in stock prices before and after the release of information, in this paper, we develop a {\em non-parametric} framework and provide empirical results on it. The framework can be used to test whether the waiting times of the jumps detected from the intraday data are abnormally distributed around a specific type of announcement compared with general jump dynamics. In particular, the waiting time can be i)``forward waiting time from the arrival of news to the nearest jump that follows and ii) ``backward waiting time'' from the nearest jump that precedes the arrival of news. Forward waiting times can be used to analyze {\em markets' post-event reactions} to understand how fast the markets react to the arrival of specific types of information in terms of jumps compared with how the jumps generally arrive. Backward waiting times, in contrast, can be used to analyze {\em markets' pre-event reactions}, which may be induced by information leakage. Methodologically, the intraday seasonality of the arrival of news and jumps is a special challenge we address in the paper. 

We use this framework with extensive sets of intraday price data and company announcements at the index and equity level:  SPY intraday prices with U.S. macro announcements and intraday prices for Nasdaq Nordic large cap stocks, traded in Copenhagen, Stockholm, and Helsinki, with company announcements classified as the scheduled and non-scheduled news. Data from Nasdaq Nordic is intriguing because all three markets are based on the same exchange rules, which makes the comparison of results for different markets interesting. Additionally, we have access to an extensive number of news events that are classified as scheduled and non-scheduled announcements, which is crucially important to verify the present method. 

Although there is strong evidence of the jump dynamics around the arrival of news in financial markets \citep{Lee2008,hussain2011simultaneous,harju2011intraday,Lee2012,bradley2014analysts}, the existing literature does not sufficiently elaborate the statistical association between the arrival of different types of news and jumps. An important exception is \citep{Lee2012} that investigates the predictability of jumps by using different information variables with U.S. data. Moreover, \citet{bradley2014analysts} examine how analyst recommendation releases are related to detected jumps. 
The main difference between our paper and \citep{Lee2012} and \cite{bradley2014analysts} lies in the objectives and methodologies: \cite{Lee2012} and \cite{bradley2014analysts} use logistic regression to predict jumps by using information about macroeconomic and firm-specific jump predictors (i.e., information variables), whereas we provide a {\em non-parametric framework} for analyzing the {\em statistical properties} of the forward and backward waiting times of the jumps. The regression method used in 
\cite{Lee2012} and \cite{bradley2014analysts} is applicable if the focus is to predict short-term jumps with a multivariate test using multiple types of announcement releases simultaneously or to distinguish systematic jumps from idiosyncratic jumps.\footnote{Moreover, recently stock price jumps have been predicted using tractable models as well as machine learning methods with limit order book data \citep{Cont2010,Kercheval2015a,ntakaris2017benchmark,tsantekidis2017using,tsantekidis2017forecasting,passalis2017time,tran2018temporal,sirignano2018universal}.} However, the method presented in this paper has advantages for studying the {\em properties} of the jump dynamics around the arrival of news rather than just predicting jumps. First, this method can be used to analyze not only immediate market reactions but also delayed reactions and possible pre-reactions to the forthcoming arrival of information in terms of jumps, even several days ahead. Second, the advantage of our approach is that it can be used with a small number of announcement events, even with a single announcement, whereas regression methods require larger sample sizes. Third, this method is non-parametric and does not require additional model assumptions except those made in detecting jumps. Fourth, the present approach can be used with multiple stocks assets (stocks) at the same time as it automatically addresses the fact that different assets may have different numbers of jumps due to market liquidity or other reasons. Finally, jumps and announcements may have strong intraday seasonality. According to the data sets we use, most of the jumps are mostly concentrated in the first trading hour. In contrast, the arrival times of the announcements can be differently distributed depending on the type of the release.\footnote{For example, our data show that scheduled company announcements mostly arrive before and at the opening time and then at noon, whereas non-scheduled announcements are more evenly distributed. Additionally, a type of macro announcement can systematically arrive at a fixed hour, for example, at 8:30 a.m.} Consequently, intraday seasonal patterns must be taken into account to estimate the actual contribution of the arrival of announcements to the arrival of jumps; otherwise, we could accidentally associate economically unimportant announcements with jump dynamics, especially if they arrive during a period when the jump activity is high for completely different reasons.  

%\cite{Lahaye2011} provided three sources of jumps associated with news: (1) selected important news, (2) foreign news and news out of list, and (3) idiosyncratic liquidity shocks from traders moving into and out of markets. Additionally, the author provided empirical evidence that co-jumps are generated mainly by macroeconomic announcements. \cite{Lee2012} also found that jump arrivals are predictable and normally distributed after macroeconomic information releases, but she did not consider the {\em sizes} of jumps. \cite{Miao2014} examined the association between macroeconomic news arrivals with S\&P500 futures. They documented that most jumps are detected in the first five minutes of a trading day. \cite{Boudt2014} investigated liquidity and news releases around jumps based on stocks in the Dow Jones Industrial Average Index. One interesting finding is that the number of trades is a key driver of jumps. \cite{Bradley2014} studied the effect of an analyst’s recommendation and found that markets react to these recommendations significantly in terms of jumps. An analyst’s advice is still the most important information source for investors, even though the distribution timing is delayed by 30 minutes on average.

This paper is also related to the massive event study literature that examines returns around the arrival of news \citep[see, for example,][and references therein]{corrado2011event,oxley2009arms}. Importantly, the present and the conventional event study methods examine different questions: Whereas the conventional method can be used to analyze gradual reactions \citep[see][]{velasquez2018layoff}, the present framework is designed to examine sudden, strong reactions (i.e., jumps) in markets that are induced by the non-gradual arrival of information. Particularly, jumps happen if many investors decide to trade in a given direction at the same time, which can be thought to indicate that several investors receive information at the same time in a non-gradual, channeled way. Methodologically, the conventional event studies do not separate the diffusion component from the jump component, and moreover, typically, they do not capture time-varying variance.\footnote{In addition to equities and stock market indexes, the proposed approach is applicable for any type of asset for which price observations are available, including Foreign Exchange \citep{Bates1996} or commodity prices \citep{schwartz1997stochastic} and even real investments \citep{dixit1994investment,kanniainen2009can} as long as their value processes can be made observable.}

The paper is organized as follows. In Section \ref{Sec:Method}, we introduce the framework. Section \ref{Sec:Data} describes the data used in this paper and in Section \ref{Sec:Empirical_Analysis}, we demonstrate our framework empirically and provide empirical results. Finally, in Section \ref{Sec:Conclusion} concludes.

%% METHOD
\section{Framework} \label{Sec:Method}

This section aims to develop a procedure for investigating whether certain announcements can be statistically associated with detected jumps\footnote
{With respect to jump detection methods, \citep[see][and references therein]{{huang2005relative,barndorff2006econometrics,andersen2007no,jiang2008testing,Lee2008,corsi2010threshold,lee2010detecting,ait2011testing}}.} that precede or follow the announcements. In this paper, the association between news events and detected jumps is studied with a measure of the {\em time distance} between a detected jump and an announcement. The distance can be in either forward or backward; forward distances can be used to assess how fast new information is adapted in stock prices, and backward distances can be used to evaluate markets' pre-reactions (due to possible information leakage). Backward and forward time distances between announcements and detected jumps are also called ``waiting time'' in this paper\footnote{The backward (forward) time distance can be considered the waiting time from a detected jump (an announcement) to an announcement (a detected jump). The terms ``time distance and ``waiting time'' are used interchangeably in this paper.}. 

The shorter the forward time distance is for  a given set of announcements, the faster the markets adapt to an information shock by the informed investors. One could also suggest that forward time distances can be used to measure how important an announcement is to the financial markets. If a set of announcements do not convey much new important information to investors, then no extra jumps are generated by the arrival of the announcement, and the price changes can be explained by pure diffusion, in which case the forward time distances from an announcement on the next jump can even be days.

The interpretation of backward distances, however, can be quite different. Stock prices could pre-jump to forthcoming announcements even some days ahead due to non-gradual (semi-)public or ``channeled'' information leakage so that the leakage efficiently reduces the information asymmetry among the market participants, in which case there will not necessarily be other jumps in the following days due to that announcement. Therefore, with backward distances the question is not whether announcements follow jumps just immediately but instead whether the backward distances are abnormally short or abnormally long. Overall, backward distances can be used to analyze i) potential information leakage with non-scheduled news whose timing should not be predictable or ii) how markets pre-process information about forthcoming scheduled news with predictable arrival times (but with non-predictable contents). 

To investigate the statistical association between the given set of announcements and detected neighbor jumps, we introduce and implement a nonparametric statistical framework. With the framework, the distribution of the empirical waiting times around announcements is compared to the distribution on the reference data where waiting times are randomized by taking intraday seasonality patterns into account. 
%Generally speaking, the average value of the observed waiting times around the actual (empirical) news arrivals is lower than the average value obtained with the reference data, then the announcements, on average, can be said to have abnormally high contribution to (detected) jumps. 
In this section, we first define backward and forward distances (waiting times) and introduce how they are measured in terms of trading time. Then, a method for generating the reference data is introduced.

\subsection{Definitions of forward and backward time distances}

Our methodology is based on i) the observed forward time distance between an announcement and the first detected jump that follows and on ii) the observed backward time distance between an announcement and the latest detected jump that precedes the announcement. We hereafter denote these forward and backward temporal distances, namely, waiting times, by $d^+$ and $d^-$, respectively. Generally, $d$ without a superscript refers to $d^+$ and $d^-$.

To introduce the methodology, let $t_{i,k}$ be $k$'th announcement time-stamp associated with asset $i$. Moreover, at each time interval $[s_{i, 1}, s_{i, 2}), [s_{i, 2}, s_{i, 3}), \dots$, we apply a jump detection method, where $s_{i,m}$ refers to the $m$th time-stamp for stock $i$. The set of the beginning points of the intervals with the detected jumps is $\mathcal{T}_i$; that is, there is a jump detected in $[s_{i, m}, s_{i, m+1})$ if and only if $s_{i,m} \in \mathcal{T}_i$. Moreover, let $\mathcal{T}_{i,k}^+$ be the set of beginning points of jump intervals that end after the arrival of the $k$th announcement and $\mathcal{T}_{i,k}^-$ be the set of beginning points of jump intervals that end no later than the arrival of the $k$th announcement. More formally, 
\begin{itemize}
\item [-] $\mathcal{T}_{i,k}^+ \cup \mathcal{T}_{i,k}^- = \mathcal{T}_{i}$ and $\mathcal{T}_{i,k}^+ \cap \mathcal{T}_{i,k}^- = \O$,
\item [-] and given that $s_{i,m} \in \mathcal{T}_{i}$, $s_{i,m} \in \mathcal{T}_{i,k}^+$ if and only if $s_{i, m+1} > t_{i,k}$, otherwise $s_{i,m} \in \mathcal{T}_{i,k}^-$.
\end{itemize}

Let $s_{i,h} \in \mathcal{T}_{i,k}^+$ be the beginning point of the {\em nearest} interval {\em with a detected jump} that {\em follows} the $k$th announcement; that is, $s_{i,h} = \min\left(\mathcal{T}_{i,k}^+\right)$. By the definition, this jump interval $[s_{i,h}, s_{i,h+1})$ ends after announcement time-stamp $t_k$, but can begin before $t_k$. The waiting time from the announcement time-stamp $t_{i,k}$ to the beginning of the jump interval $s_{i, h}$ is referred as the forward distance and is defined as follows:
\begin{equation}\label{EQ:d+}
d_{i,k}^+ = \max\left(s_{i,h} - t_{i,k}, 0 \right).
\end{equation}
If the jump interval strictly follows the announcement time-stamp, that is, $s_{i, h} > t_{i,k}$, then $d_{i,k}^+ > 0$. However, if the announcement time-stamp is within the associated jump interval, that is, $s_{i, h} \leq t_{i,k} < s_{i, h+1}$, then $d_{i,k}^+ = 0$.

Now, let $s_{i,h} \in \mathcal{T}_{i,k}^-$ be the beginning point of the {\em nearest} interval {\em with a detected jump} that {\em precedes} the $k$th announcement; that is, $s_{i,h} = \max\left(\mathcal{T}_{i,k}^-\right)$. By the definition, this jump interval $[s_{i,h}, s_{i,h+1})$ begins before the announcement time-stamp and cannot end after the announcement time-stamp $t_k$. The waiting time from the end of the jump interval $s_{i, h+1}$ to the announcement time-stamp $t_k$ is referred to as the backward distance and is defined as follows:
\begin{equation}\label{EQ:d-}
d_{i,k}^- = t_{i,k} - s_{i,h+1}.
\end{equation}
The backward distance is always non-negative because, by the definition, $t_{i,k} \geq s_{i,h+1}$ if (and only if) $s_{i,h} \in \mathcal{T}_{i,k}^-$. 

\begin{figure}[!h]
\begin{center}
\includegraphics[width=0.6 \textwidth, angle=0]{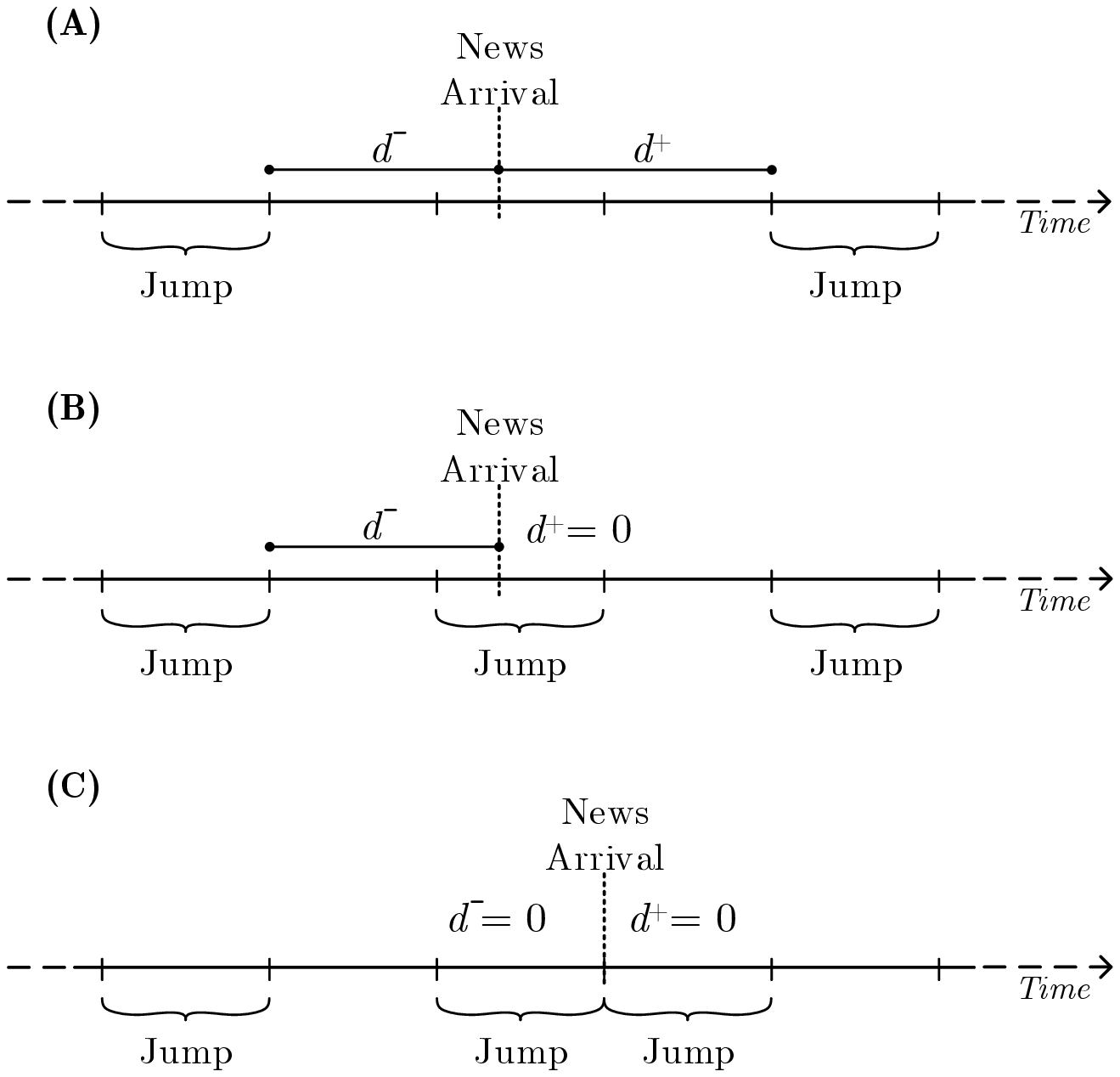}
\caption{\footnotesize Plot A demonstrates the determination of forward and backward distances, $d^+$ and $d^-$, when the distances to the nearest jump periods are greater than zero. In plot B, the announcement time-stamp is within a jump interval, and thus, $d^+ = 0$ (see Eq. \ref{EQ:d+}), whereas $d^-$ is greater than zero. Plot C demonstrates a hypothetical case, in which the arrival of the announcement takes places {\em exactly} between two associated subsequent jump intervals, in which case, $d^+ = 0$ and $d^- = 0$. \label{FIG:distanceDemonstration}}
\end{center}
\end{figure}

Importantly, forward distances are defined in such a way that the announcement time-stamp can be within the associated jump interval; that is, $s_{i, h} \leq t_{i,k} < s_{i, h+1}$. In this special case (where $d_{i,k}^+  = 0$), we know only that there was a jump within the same interval with the announcement arrival but not whether the actual jump took place exactly before or after the announcement time-stamp.\footnote{Notice that the jump detection methods determine  {\em intervals} within which there are jumps with a given confidence level in the stock price rather than exact jump time-stamps.} Therefore, forward distance measures markets' reaction (in terms of jumps) that has taken place during an interval after or at the arrival of an announcement. This and other cases are demonstrated in Figure \ref{FIG:distanceDemonstration}.

However, backward distances are defined such that the announcement time $t_k$ does not belong in the associated half-closed interval $[s_{i,h}, s_{i,h+1})$. Therefore, regarding backward distances, we know that the actual jump has not taken place after the arrival of the announcement, and therefore, the backward distance cannot accidentally measure the market's post-reactions. This feature is very important if backward distances are used to examine markets' pre-reactions to information arrivals. 

In this paper, we i) measure the time distances (waiting times) in terms of trading time and ii) include both trading and non-trading time announcements. The time distances could also measured in terms or calendar time and/or the non-trading time announcements could be excluded. Results for these settings are available upon request, but in this paper we report results only for the baseline settings (waiting times measured by trading time and both trading and non-trading time announcements are included), which is preferred for a number of reasons. First, in terms of sample size, It is a clear advantage to include announcements that arrive during non-trading hours. For example, for the scheduled announcements of the Danish companies in our data set, 338 observations are available, but only 156 of them are published during trading hours. Second, the time distances measured in terms of trading time have advantages. In particular, the use of actual (calendar) hours with non-trading time announcements can be problematic for measuring the market's reaction time as the lower bound of the market's reaction time can be several hours. For example, if an announcement is released in the late afternoon only a few minutes before the closing time, say at 3:58 p.m., and the markets have no time to react to the announcement during the same day, the announcement is then reacted to the following morning around opening time, say at or slightly after 9:30 a.m., in which case the calendar time distance cannot be less than 17.5 hours. Therefore, our approach that measures time distances in terms of trading time considers last-minute cases the same as announcements published around the opening time. It is good to keep in mind that several announcements that arrive during non-trading hours in the same night are all equally associated with the next jump, which, can be an issue if the number of announcements arrived during the same night is large.

%\begin{table}%[htbp]
%\caption{Approaches for measuring waiting time distances. \label{Tab:ApprDistances} }
%  \centering
%  %\scalebox{0.85}
%	{\small
%\begin{tabular}{lll}
%\\
%{\bf Approach} & {\bf Non-trading time} & {\bf Trading/calendar} \\
%    {\bf } & {\bf announcements} & {\bf time distances} \\
%\hline
%         I &   Included & Trading time distances \\
%         II &   Excluded & Trading time distances \\
%         III &   Included & Calendar time distances \\
%         IV &   Excluded & Calendar time distances \\
%\hline
%\end{tabular}  
%	}
%\end{table}

Suppose that there is a non-trading period between an announcement and a jump such that news arrival time is $t$, a related jump interval begins at $s$, a non-trading period begins at $\tau_1$ and ends at $\tau_2$, where $t < \tau_1 < \tau_2 < s$, and the length of the jump interval is $\Delta s$. Instead of using
\[
d^+ = s-t = (s-\tau_2) + (\tau_2-\tau_1) + (\tau_1 - t),
\]
we take the minimum between  the length of the non-trading period and the length of the jump interval
\[
d^+ = (s-\tau_2) + \min(\tau_2-\tau_1, \Delta s) + (\tau_1 - t).
\]
In this way, the non-trading period (in US markets 4:00pm - 9:30am) is never measured to be longer than the length of the jump period (15 minutes). Moreover, if an announcement arrives during non-trading hours, that is, $\tau_1 < t < \tau_2 < s$, then the forward distance is defined as
\[
d^+ = (s - \tau_2) + \min(\tau_2 - t, \Delta s).
\]
That is, in this case, the distance from the announcement to the opening time is never longer than the length of the jump period. The backward distances are calculated analogously.

\subsection{Generation of reference data samples}\label{subsubsec:data_simulations}

\subsubsection{Idea}
The present framework can be used to answer the following question: Given that there is certain intraday seasonality in jumps and the arrival of news, how do jumps arrive around announcements compared with how the jumps generally arrive? To answer this question, a {\em reference data sample} is generated to isolate the impact of announcements on jumps from the impact of intraday seasonality patterns. In particular, the main idea is to compare distributions of i) the empirical forward and backward distances between the actual arrival of news and the detected jump intervals, $d^+, d^-$ and ii) the forward and backward distances between generated reference time-stamps and detected (empirical) jumps, $\tilde{d}^+, \tilde{d}^-$, respectively. A central question here is how the reference time-stamps should be generated to capture the intraday seasonality of the arrival of news and jumps, which strongly appear in the data as later demonstrated in Figures~\ref{FIG:announcement_histograms}  and \ref{FIG:jump_histogram}, respectively \citep[see also][]{Lee2008,Lee2012}. As Figure \ref{FIG:jump_histogram} demonstrates, the jumps are mostly concentrated in the first trading hour in all the markets (Copenhagen, Stockholm, and Helsinki). 

Different types of announcements can have different intraday seasonal patterns, and this difference must be taken into account in the generation of reference data samples, which we address by generating the reference data sets with an empirical distribution of the arrival times of the announcements. In this paper, time-stamps associated with a specific asset (company) in the reference data sets are randomly generated from the same asset's empirical distribution for the time-stamps of the arrival of the announcements. If there is variation in the clock times of the arrival of the announcements, that is, the given announcements do not always arrive at certain clock times, one may use kernel density estimation for data smoothing. In this regard, we follow \cite[][]{botev2010kernel} to optimize the bandwidth for irregularly arriving (company) announcements\footnote{Their Matlab package is available on the MathWorks webpage \url{www.mathworks.com/matlabcentral/fileexchange/14034-kernel-density-estimator}}. When the arrival of the announcement is fixed to a specific time-stamp(s), one may apply an empirical distribution directly, as we do with some macroeconomic announcements. 

To justify our approach to use an empirical distribution to generate intraday time-stamps (instead of a uniform distribution, for example), consider a hypothetical example in which the arrival times of specific announcements over multiple days are mostly concentrated in the first trading hour (data set A) whereas the other set of announcements, arriving on multiple days, typically occurs at around noon (data set B). In contrast, suppose there is a strong intraday seasonal pattern in the arrival of the jumps, too. In particular, most jumps typically occur during the first trading hour, which is intuitive as then the markets react to the news that arrived after the closing time, whereas jumps around noon are clearly more uncommon. Consequently, if the reference data samples are generated from the same distribution (e.g., uniform distribution) for announcement sets A and B without taking the announcements' actual and different seasonal patterns into account, then the contribution of the announcements to the arrival of the jumps is likely to be overestimated with data set A and underestimated with data set B.

\subsubsection{Procedure}

We use a notation where the trading days (observed from the empirical data) are denoted for asset $i$, $i = 1, 2, \dots, N$, by $\{T_{i, 1}, T_{i, 2}, \dots, T_{i, m_i}\}$, where $m_i$ is equal to the numbers of asset $i$'s trading days in the data. $T_{i, j}$ can be considered an integer, which represents the beginning (i.e., midnight, 00:00 a.m.) of day $j$  (ref. \texttt{datenum} in \texttt{Matlab}). Given that the length of one calendar day is one, then the specific clock times on specific dates could be denoted by a floating number; for example, $T_{i, j} + 0.75$ would be day $j$ at 6 p.m. In addition, for asset $i$, series $\{n_{i,1}, n_{i,2}, \dots, n_{i,m_i}\}$ denotes the number of announcements that are associated with days $\{T_{i, 1}, T_{i, 2}, \dots, T_{i, m_i}\}$, $j = 1, 2, \dots, m_i$; that is, $n_{i,j}$ denotes the number of announcements related to asset $i$ on the $j$th trading day.

A reference sample is generated as follows:
\begin{enumerate}
	\item In the first step, for asset $i$, we generate series $\{n_{i,1}, n_{i,2}, \dots, n_{i,m_i}\}$ by simulating $n_i$ draws from the uniform distribution to be associated with days $\{T_{i, 1}, T_{i, 2}, \dots, T_{i, m_i}\}$. Here, $n_{i, j}$ corresponds to the number of time-stamps on the $j$th day for asset $i$ in the {\em reference data set}. That is, if there are $k$ realizations for $T_{i,j}$, then we set $n_{i,j} = k$. The total number of time-stamps generated for asset $i$ equals the number of actual announcements for the asset, $n_i$, observed for the empirical data, and thus, $\sum_{j=1}^{m_i} n_{i,j} = n_i$.  $n_i$  is the total number of announcements using trading and non-trading hours.

	\item In the second step, for asset $i$ on day $T_{i, j}$, we generate $n_{i, j}$ time-stamps, $\{\tau_{i, j, 1},$ $\tau_{i, j, 2}, \dots, \tau_{i, j, n_{i,j}}\}$, with an {\em empirical distribution} to take the intraday seasonality into account. The empirical distribution can be asset-specific, or alternatively, one can estimate it using aggregated data over all the assets in the sample.\footnote{We implement both approaches in our empirical demonstration}. Here, kernel density estimation can be applied to smooth data, or alternatively, if the arrival times are clearly fixed to certain clock times, then an unsmoothed empirical distribution can be used instead.
	
	Here, $0 \leq \tau_{i, j, k} \leq 1$ for $i = 1, 2, \dots, N$ $j = 1, 2, \dots, m_i$, and $k = 1, 2, \dots, n_{i,j}$. Moreover, as specified above, $T_{i, j}$ is 00:00 a.m. $T_{i, j} + 1$ is 24:00 p.m. on the $j$th day, and therefore, given that $n_{i,j} > 0$, the generated time-stamps for the $i$th asset and the $j$th day are $\{T_{i, j} + \tau_{i, j, 1}, T_{i, j} + \tau_{i, j, 2}, \dots, T_{i, j} + \tau_{i, j, n_{i,j}}\}$. We repeat this step for all the trading days with positive $n_{i,j}$, $j = 1, 2, \dots, m_i$.
	\item Third, we repeat steps 1 and 2 for all $N$ assets.
	
	\item Then we apply the data processing described in Section \ref{subsubsec:time_distances}. In particular, we make i) the distances from the closing time to the arrival of news, ii) the distances from the arrival of the news to the opening time, and iii) the length of the non-trading periods to equal to no more than the distance of the jump period, after which the distances with the empirical and reference data sets are measured in terms of the trading hours. 
	
	\item Finally, steps 1--4 can be iterated to make multiple reference samples. 
\end{enumerate}

\subsubsection{Methods for statistical analysis}\label{SUBSECT:Methods}
Given that announcements provide valuable information to financial markets, the forward distances can be expected to be abnormally short compared with what they generally are. Therefore, we may postulate that $\mathbb{E}(d^+) < \mathbb{E}(\tilde{d^+})$, where $d^+$ refers to the distances between the empirical announcement times and the detected jumps and the $\tilde{d^+}$ distances between the generated time-stamps and the detected (empirical) jumps. Backward distances that reflect possible information leakages can methodologically be more complex. It is clear that if information leakage takes place just before the actual time of an announcement, then the associated backward distance can be abnormally short. However, what if the information leakage takes place two days before the actual time of the public announcement while the waiting time for the jumps generally is one day? In this case, the stock price could jump two days before the actual time of the announcement, and moreover, if the leakage efficiently reduces the information asymmetry among the market participants and there are no other important news releases, then there will be no other jumps in the following days. This reasoning suggests that especially with backward distances it is important to independently test whether the empirical backward distances are, on average, smaller or larger than the reference ones. Consequently, we test the null hypothesis against two alternative hypotheses: i) $\mathbb{E}(d^-) < \mathbb{E}(\tilde{d^-})$ and ii) $\mathbb{E}(d^-) > \mathbb{E}(\tilde{d^-})$.

Two-sample can be used to test the empirical distances against the generated reference distances to analyze the contribution of a set of announcements to the arrival of the jumps. The calculation of means is, of course, trivial, but the associated p-values are more demanding, because in this case the distributions are asymmetric, the variances are unequal, and the sample sizes are different. Under these circumstances, Welch's U-test, often called an unequal variances t-test for ranked data, can be applied for the means. We evaluate whether the forward and backward distances specified in Eq.(\ref{EQ:d+}) and (\ref{EQ:d-}), can be said to be statistically smaller or larger between the empirical and generated reference time-stamp data sets. Regarding the size of the reference sample for goodness-of-fit tests, \citet{Bera2013} suggest that in a two-sample test the reference sample should be larger than the sample to be examined, and the authors find that satisfactory results are obtained with two-sample tests when a simple rule of thumb that the number of observations in the reference sample equals the squared number of observations in the test sample is used. Therefore, in our context, this would suggest that the data generation for the reference distribution should be iterated as many times as the number of empirical announcements observed in the data. Because the size of an empirical sample is $n = \sum_{i=1}^N n_i$, where $n_i$ is the number of actual (empirical) announcements of asset $i$ and $N$ is the number of assets, we should generate $n$ individual copies of the reference data sample (each with different random seeds). One may also want to compare the sample medians. However, the p-values for the medians are more complex, but one can always report the p-values using bootstrapping. In bootstrapping, we may generate a considerable large number, say 10,000, of reference data sets, whose sizes are the same as the size of the empirical data set. The left-sided (right-sided) p-value is then simply obtained by dividing the number of cases where the reference median is less (more) than or equal to the empirical median by the total number of reference data sets (say 10,000). The same procedure can be used for bootstrapped p-values for the means.\footnote{Alternatively, instead of focusing on the first moment, one may also want to run the Kolmogorov-Smirnov test, or other global non-parametric tests for homogeneity, can be used to decide whether two random samples have the same statistical distribution over the entire domain. We have computed results for one-sided two-sample Kolmogorov-Smirnov tests, which are available upon request.} It is also possible to examine even single events with the present framework non-parametrically. One can test a single event against the generated reference distribution and non-parametrically calculate the associated p-values. In this paper, we do not empirically demonstrate the calculation of the p-values for specific events, but nevertheless, it is a clear advantage that our methodology can be used with a very small set of announcement events.

\subsection{Jump detection method used}

In this paper, we use the framework by applying the existing jump detection method introduced in \citep{Lee2008}, which is one of the most referred methods in the recent literature, but other methods could be used, such as \citep{huang2005relative,andersen2007no,corsi2010threshold,lee2010detecting,ait2011testing}. By following \cite{Lee2008}, we assume the dynamic of the asset $i$ log-returns follows the stochastic differential equation
\[
d\log S_i(t) = \mu_i(t)dt + \sigma_i(t)dB_k(t)+ Y_i(t)dJ_i(t),
\]
where $B_i(t)$ is a Brownian motion, which is independent of the jump component $\int_{0}^{t}Y_i(s)dJ_i(s)$, $Y_i(t)$ is the jump size, and $J_i(t)$ is a counting process independent of $B(t)$. It may be a non-homogeneous Poisson-type jump process, and therefore, as \citet{Lee2008} argue, ``scheduled (deterministic) events such as earnings announcements are allowed to affect jump intensity''. This specification incorporates a sufficiently large class of asset price dynamic setting, such as stochastic volatility models in \citep{heston1993closed} and \citep{schobel1999stochastic} plus the finite activity jump semi-martingale class in \citep{Barndorff-Nielsen2004}.

The jump detection statistic of \citep{Lee2008} is
\begin{equation}\label{EQ:L}
\mathcal{L}_i(k) \equiv \dfrac{\log [S_i(t_k)/S_i(t_{k-1})]}{\widehat{\sigma}_i(t_k)},
\end{equation}
where $\widehat{\sigma}_i(t_k)$ is the realized bipower variation for asset $i$ at time $t_k$, which is used for the instantaneous volatility estimation to make the technique robust to the presence of jumps in previous time intervals. 

Intuitively, if there is no jump within the interval $k$, then the size of $\mathcal{L}(k)$ should be significantly smaller than if there were a jump. \cite{Lee2008} obtain the following rejection region for the null hypothesis (of no jump within the $k$th time interval):
\[
\frac{|\mathcal{L}_i(k)| - C_n}{S_n} > -\log(-\log(1-\alpha)),
\] 
where 
\[ 
C_n = \frac{(2\log n )^{1/2}}{\sqrt{2/\pi}} - \frac{\log \pi + \log(\log n)}{2\sqrt{2/\pi}(2\log n )^{1/2}}, \text{\quad}
S_n = \frac{1}{\sqrt{2/\pi} (2 \log n)^{1/2}}, 
\]
$n$ is the sample size and $\alpha$ is the level of significance.

% DATA
\section{Data} \label{Sec:Data}
To empirically demonstrate the present framework, we use two data sets. First, we use U.S. macro announcements with intraday observations for SPY, the exchange-traded funds (ETFs) of the S\&P500 Index.   Second, we use scheduled and non-scheduled {\em corporate} announcements with intraday stock prices from Nasdaq Nordic from 2 January 2006 to 31 December 2009. In particular, we analyze three sets of stocks separately: i) 20 large-cap Danish companies traded on the Copenhagen exchange, ii) 28 large-cap Swedish companies traded on the Stockholm exchange, and iii) 29 large-cap Finnish companies traded on the Helsinki exchange. 
 
\subsection{Macro announcements and SPY data}

We analyze how the S\&P500 Index, or SPY, which its ETFs are traded between 9:30 a.m. and 4:00 p.m., react to U.S. macro announcements between January 2001 and December 2013. The U.S. macroeconomic announcements are from the Bloomberg World Economic Calendar, and to limit the length of the paper, we focused on the following announcements: ADP Employment Change, CPI Core Index SA, Change in Nonfarm Payrolls, Chicago Purchasing Manager, FOMC Rate Decision (Upper Bound), Factory Orders, Initial Jobless Claims, Nonfarm Productivity, and the Underemployment Rate. To detect jumps in the index ETFs, we use the mid-prices of the SPY prices, which are available with millisecond precision time-stamps. After cleaning, we sample the prices at a 15-minute frequency as we do with equity data from Nasdaq Nordic.\footnote{We thank Aarhus University for providing the data for the macro announcements and SPY during Ye Yue's visiting scholar period.} 

\begin{table}[!h]
	\caption{The numbers and arrival times of U.S. macro announcements analyzed in this paper between January 2001 and December 2013. \label{Tab:numberOfMacroAnnouncements} }
	\ \\
	\centering
	\resizebox{1\textwidth}{!}
	{
		\begin{tabular}{l|ccc|c}
			& Trading time & Non-trading time & Total & Arrival \\
			& announcements & announcements &       & time(s) \\
			\hline
			ADP Employment Change & 0     & 89    & 89    & 8:15 AM \\
			CPI Core Index SA & 0     & 90    & 90    & 8:30 AM \\
			Change in Nonfarm Payrolls & 0     & 151   & 151   & 8:30 AM \\
			Chicago Purchasing Manager & 154   & 0     & 154   & 09:45 AM, 10:00 AM \\
			FOMC Rate Decision (Upper Bound) & 104   & 3     & 107   &  7:00 AM, 8:20 AM, 10:55 AM, \\
			&       &       &       & 12:30 PM,  2:00 PM, 2:12 -- 2:20 PM \\
			Factory Orders & 152   & 0     & 152   & 10:00 AM, 03:00 PM \\
			Initial Jobless Claims & 0     & 668   & 668   & 08:30 AM, 10:30 PM \\
			Nonfarm Productivity & 0     & 103   & 103   & 8:30 AM \\
			Underemployment Rate & 0     & 28    & 28    & 8:30 AM \\
		\end{tabular}%
	}
\end{table}

In Table \ref{Tab:numberOfMacroAnnouncements}, the macro-announcement data used in this paper is summarized. Most of the announcements have a specific release time, typically 8:30 a.m., which is before the opening time of an exchange (9:30 a.m.). Moreover, three announcement types (Chicago Purchasing Manager, Factory Orders, and Initial Jobless Claims) had two possible release times. For announcements that have no more than two release times, we do not use kernel density estimation; the reference data sample is based on the exact clock times. The FOMC Rate Decision is the only announcement that has some variation in release times, and thus, kernel density estimation is applied to this announcement. The table also shows that six out of nine announcement types arrive during non-trading hours only. 

Table \ref{Tab:numberOfMacroAnnouncements} summarizes the macro announcement data. It is observed that there are strong overlaps in the arrivals of macro announcements. The majority of selected macro announcements are released at 8:30 a.m.—an hour before the opening time of an exchange (9:30 a.m.). Additionally, one announcement type can have several announcement times. For example, Chicago Purchasing Manager, Factory Orders, and Initial Jobless Claims have two possible release times, and FOMC Rate Decision has multiple release times. Moreover, Initial Jobless Claims, Nonfarm Productivity, and Underemployment Rate are announced completely outside of trading hours.

From the detected jumps, the jump intensity of SPY is quite lower than that of equities in Nasdaq Nordic.  The jump rate (the number of jumps divided by the number of trading days) for SPY is only 0.12. In contrast, the corresponding rates are between 0.18 and 0.51 across stocks on the Copenhagen exchange, between 0.22 and 1.90 across stocks on the Stockholm exchange, and between 0.14 and 0.84 across stocks on the Helsinki exchange in the samples. This might be due in part to the difference in liquidity of the markets. Another potential explanation is that the influence of announcements between Nordic firm-level news and U.S. macroeconomic news is irreversible. U.S. macro news affects Nordic markets significantly; however, Nordic firm-level news does not affect the U.S. market. More discussion on the jumps of equities and indexes can be found in \cite{Lee2008}.
Moreover, jumps are found to arrive with daily seasonality:  89\% of the jumps occur during the first half-hour (9:30–10:00 a.m.), and 92\% occur during the first hour (9:30–10:30 a.m.). 

\subsection{Company announcements and stock price data from Nasdaq Nordic}

In Nasdaq Nordic, the market places in Stockholm, Copenhagen, and Helsinki are based on the same exchange rules, which makes the comparison of results for different markets interesting. For these market places, we have access to an extensive number of news events that are classified as scheduled and non-scheduled announcements. This classification is important to verify our methodology; if we get i) statistically insignificant results for backward distances with non-scheduled announcements (whose arrival should be non-predictable by the definition) and ii) statistically significant results for forward distances, the methodology is verified to be able to accept and reject the null hypothesis that the distances come from populations with the same distribution. Regarding the backward distances with non-scheduled announcements, it is, of course, always possible that there is information leakage to some extent in the real markets, in which case our methodology cannot be said to be invalid even if the null hypothesis is rejected in this particular case. To put it another way, given that the methodology is verified to be able to accept and reject null hypotheses, if the null hypothesis is accepted (rejected) for backward distances with non-scheduled news, we get evidence that there was no information leakage (has been information leakage) regarding the examined set of news. 

We use firm-specific announcements delivered by Nasdaq Nordic, which continuously publishes first-hand announcements delivered by listed companies.\footnote{\url{http://www.nasdaqomxnordic.com/news/companynews}; see the page for detailed information.} These include, for example, earning announcements, news about an acquisition, take-over bid, capital increase, new product launch, expansion into new markets, signing of alliances, etc.\footnote{Announcements provided by Nasdaq cover the messages that were filed with Nasdaq  by the respective companies. Each company may publish additional, non-regulatory news on their own website, which are not part of our data samples.} Nasdaq associates each announcement with an exact time-stamp and a company name, which we then match with independent international securities identification number (ISIN) codes. For example, ``Finnair sells one Embraer 170 aircraft'' is announced at 31.12.2010, 08:45 AM under the category ``Company Announcement'' and is associated with ``Finnair Oyj''. 
	
Nasdaq Nordic Data provides a classification with different categories. However, in this paper we do not restrict our study to a specific news class, such as earnings announcements, as many other studies do; instead, as in \citep{siikanen2017liquidity}, the announcements are re-categorized into two specific groups, scheduled and non-scheduled announcements, in order to demonstrate the methodology with announcements whose arrival time is predictable and non-predictable. An announcement is classified as scheduled if its exact publishing date is known to the public beforehand. This happens if the date is given in advance in earlier stock exchange releases or on the financial calendar. Correspondingly, an announcement is classified as non-scheduled if external stakeholders do not know when the announcement will come. In particular, a release is considered non-scheduled if it is irregular, its publishing schedule is not given and cannot be reliably estimated, or the release is obviously unexpected. To be on the safe side, announcements whose publishing time span is given non-specifically in earlier stock exchange releases or that are somewhat regular by nature, such as proposals to annual general meetings by the board or nomination committee, are excluded. Most of the excluded announcements are notices to convene annual general meetings, notices of the publication of annual reports or summaries, and invitations to press conferences related to publishing financial reports. Moreover, announcements that clearly contain no new information are excluded. In the Nordic markets, announcements can be released twice in the local language and in English at slightly different times, in which case only the first time-stamp is applied. 

\begin{figure}[!h]
\begin{center}
\includegraphics[width=0.8\textwidth, angle=0]{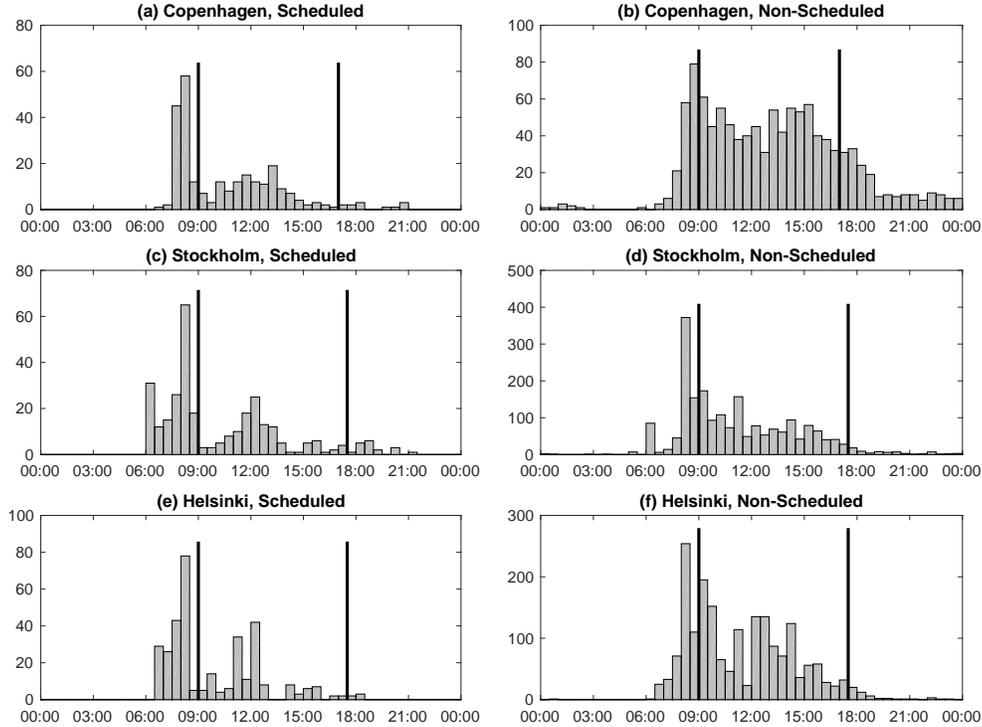}
\caption{\footnotesize Histograms of the scheduled and non-scheduled announcements for the Danish, Swedish, and Finnish large-cap companies in the samples. The vertical lines represent the opening and closing times (Copenhagen 9:00 a.m. and 5:00 p.m., Stockholm and Helsinki 9:00 a.m. and 5:30 p.m. CEST). \label{FIG:announcement_histograms}}
\end{center}
\end{figure}

Table \ref{Tab:numberOfAnnouncements} summarizes the number of scheduled and non-scheduled announcements that arrive during and after trading hours for different sets of stocks. We conducted full and filtered sets for the announcements; the full set includes all the scheduled announcements for given companies (All in the table), and the filtered set excludes announcements that had another (scheduled or non-scheduled) announcement in the neighborhood of 6 hours on both sides, respectively. That is, to eliminate days that experienced other significant events, often referred to as confounding events in the literature, for each company we include only announcements for which there were no other announcements in the neighborhood. The table shows that non-scheduled news releases largely arrive during non-trading hours, whereas scheduled news arrive mostly during trading hours. The largest announcements sample is the Swedish non-scheduled events, and the smallest sample is the Danish scheduled news. Additionally, Figure \ref{FIG:announcement_histograms} provides histograms for the arrival times of scheduled and non-scheduled announcements in three markets that include all the announcements. Scheduled announcements mostly arrive before and at opening time and then, especially on the Stockholm and Helsinki exchanges, at noon whereas non-scheduled announcements are more evenly distributed.

\begin{table}[!h]
\caption{Numbers of classified company announcements in different markets between 2 January 2006 and 31 December 2009. Scheduled refers to scheduled announcements and Non-Scheduled to non-scheduled announcements. All data includes all the announcements in the sample. Filtered data set excludes announcements that had another (scheduled or non-scheduled) announcement in the neighborhood of 6 hours on both sides. \label{Tab:numberOfAnnouncements} }
  \ \\
	\centering
  \resizebox{1\textwidth}{!}
	{
\begin{tabular}{l|ccc|ccc}

           & \multicolumn{ 3}{|c}{{\bf Scheduled announcements}} & \multicolumn{ 3}{|c}{{\bf Non-scheduled announcements}} \\

           & Trading time & Non-trading time &      Total & Trading time & Non-trading time &      Total \\

           & announcements & announcements &            & announcements & announcements &            \\
\hline
{\bf Copenhagen Large-Cap} &            &            &            &            &            &            \\

\ \ a) All data &        156 &        182 &        338 &        863 &        441 &       1304 \\

\ \ b) Filtered data set &        127 &        130 &        257 &        726 &        354 &       1080 \\

{\bf Stockholm Large-Cap} &            &            &            &            &            &            \\

\ \ a) All data &        198 &        251 &        449 &       1751 &        999 &       2750 \\

\ \ b) Filtered data set &        122 &        185 &        307 &       1295 &        759 &       2054 \\

{\bf Helsinki Large-Cap} &            &            &            &            &            &            \\

\ \ a) All data &        305 &        331 &        636 &       1799 &        749 &       2548 \\

\ \ b) Filtered data set &        152 &        186 &        338 &       1369 &        542 &       1911 \\

\end{tabular}  
	}
\end{table}

Second, the trading data we use in this paper is the tick-by-tick records of Level I order book data provided by Nasdaq Nordic.
	The total number of trading days in our data is 977 out of the four years (January 2006--December 2009). 
Before the middle prices are calculated for each stock, high-frequency data must be cleaned and preprosessed. As discussed in \citep{genccay2001introduction} and \citep{brownlees2006financial}, several typical errors in high-frequency data are caused by humans or systems. We implement the step-by-step cleaning procedures introduced in \citep[][procedures P1-P3 and Q1-Q4]{Barndorff-Nielsen2009}. Additionally, we delete all observations recorded in any trading halt interval.

After data cleaning, we adopt the following two-step procedure for data sampling:
\begin{enumerate}
\item  We sample quote records every 10 seconds to generate a regular 10-second spaced data set in time from the cleaned tick data. If there is no quotation at some 10-second time-stamp, we sample the nearest value for that time-stamp.
\item  In order to avoid a strong microstructure problem in jump detection as discussed in \citep{Lee2008} and to reduce errors in jump detection due to the use of low-frequency data \citep[see][for further details]{christensen2014fact}, we conservatively choose a 15-min sample frequency from the 10-second sample.
\end{enumerate}

\begin{figure}[!h]
\begin{center}
\includegraphics[width=1\textwidth, angle=0]{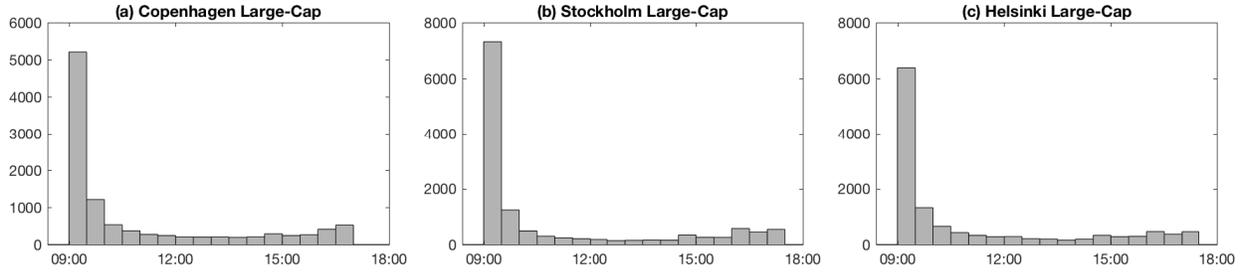}
\caption{\footnotesize The histogram of the time-stamps of the detected jumps for Danish, Swedish, and Finnish large-cap stocks. The jumps are detected using a methodology from \citep{Lee2008} with a sampling interval of 15 minutes with the 1\% significance level. The opening hours for the Copenhagen exchange are between 9:00 a.m. and 5:00 p.m. and for the Stockholm and Helsinki exchanges are between 9:00 a.m. and 5:30 p.m. \label{FIG:jump_histogram}}
\end{center}
\end{figure}

Figure \ref{FIG:jump_histogram} provides the histograms of the arrival times of the jumps with the data samples, where we apply the 1\% significance level. Clearly, the jumps are mostly concentrated in the first trading hour in all the markets. In fact, the histogram profiles are surprisingly similar across the three Nordic exchanges.

\begin{figure}[!h]
	\begin{center}
		\includegraphics[width=0.7\textwidth, angle=0]{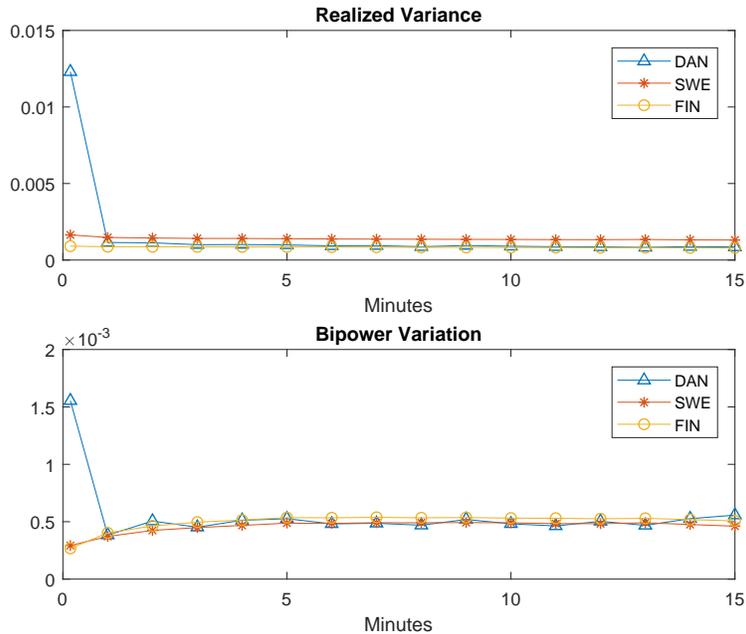}
		\caption{\footnotesize Signature plots of average realized variance and bipower variation across individual stocks in Finnish, Swedish and Danish markets. The shortest sampling interval corresponds to 10 seconds. \label{FIG:VolSignature}}
	\end{center}
\end{figure}

Compared to the US markets, Nordic markets are less liquid and therefore using signature plots of realized variance and bipower variation, we analyze what would be an appropriate sampling frequency. The signature plots are provided in Figure \ref{FIG:VolSignature}, which are calculated using 16 different sampling frequency returns: 10-second, 1-minute, 2-minute, and up to 15-minutes. The means realized variance and bipower variation are computed using prices of  20 large-cap stocks in the Danish market, 28 large-cap stocks in the Swedish market and 27 large-cap stocks in the Finnish market from 2006 to 2009. From this Figure, a strong effect of micro-structure noise on both realized variance and bipower variation can be found using 10-second returns. However the micro-structure effect gradually disappears as the low-frequency returns are sampled. This finding is confirmed in \cite{andersen1999realized} and \cite{bandi2008microstructure}. Other research suggests using relatively high frequency stock price data. For instance, \cite{liu2015does} shows that 5-minute returns are a safe choice, and \cite{zhang2005tale} argues for using all of the intraday data. Due to the illiquidity of stocks in Nordic markets, we conservatively adopted 15-minutes returns to detect jumps.

%% EMPIRICAL ANALYSIS
\section{Empirical Demonstration and Analysis}\label{Sec:Empirical_Analysis}

\subsection{S\&P500 index with macro announcements}

Table \ref{tab:Macro_means} reports the bootstrap and Welch U-test p-values for the medians and means of the forward and backward waiting times, including right- and left-tailed values. The following observations can be made:\ \\ \\
{\em Forward Distances}\\
First, we identify four macro announcement releases that are statistically associated with the forward jumps across the mean and media median tests without exception: ADP Employment Change, Chicago Purchasing Manager, FOMC Rate Decision, and Underemployment Rate. Second, the mean test provide evidence with the CPI Core Index and the Change in Nonfarm Payrolls. No test rejected the null hypotheses for the three most important macro releases, Factory Orders, Initial Jobless Claims, and Nonfarm Productivity, meaning that there is no statistical evidence for the association between these announcements and jumps in SPY. This is interesting, especially because \cite{Lee2012} provides evidence that Initial Jobless Claims releases are good jump predictors in the U.S. individual equity markets. The differences in the results may lie in the differences between the methods. Our methodology is based on the use of empirical distributions of release times, and Lee uses logistic regression to predict jumps\footnote{\cite{Lee2012} uses dummy variables for the first hours between 9:30--10:00 and 10:00--11:00, which should approximately address the intraday seasonality in the morning.} without considering the waiting times from the announcements to the jumps. In addition, there are differences in the data. We run this analysis at the market index level for SPY and not for individual stocks. Moreover, \cite{Lee2012} predicts the jumps in the next period whereas the waiting time distances in our methodology can span several periods, which makes it possible to examine slower reactions.  In fact, this can be important, because the present empirical investigations show that the jump reactions are not always immediate and a forward jump can occur a period or two late. For example, in the data for scheduled announcements on the Copenhagen exchange, there were 636 events, and in 121 cases, the forward waiting time was 15 minutes or less whereas in 227 cases the forward time was more than 15 minutes but less than half an hour. Even more importantly, backward jumps can occur several jump periods, even days, ahead, which can be addressed with the present methodology.

\ \\
{\em Backward Distances} \\
We provide some evidence of abnormal jump dynamics in the S\&P500 ETF preceding scheduled macro announcements. In particular, the medians or means for the backward distances for CPI Core Index and Underemployment Rate are statistically significantly lower for the empirical data compared with the reference data sets (statistically significant left-tailed p-values), which indicates that the distances were abnormally short around these announcements. In addition, the mean test indicate statistically significant right-tailed p-values for the FOMC Rate Decision and the median test for the Chicago Purchasing Manager, indicating that the backward distances were abnormally long around these announcements. Similarly, as with the forward distances, no statistical test suggests rejection of the null hypotheses for Factory Orders, Initial Jobless Claims, and Nonfarm Productivity, meaning that there is no statistical evidence for an association between these announcements and the preceding jump dynamics in SPY.

\begin{table}
	\caption{\textbf{Medians and means of the forward (Panel A) and backward distances (Panel B) for the macro data sample.}{\footnotesize \normalfont  ``Bootstr. p left tail'' and ``Bootstr. p right tail'' are the left- and right-tailed p-values calculated with the bootstrapping method, respectively, and the ``Welch U-test p left tail'' and ``Welch U-test p right tail'' are the left- and right-tailed p-values for the mean values calculated with the Welch U-test (unequal variances t-test for ranked data), respectively.}} 
	\label{tab:Macro_means}
	
	\centering
	\resizebox{0.95\textwidth}{!}
	{\large
		% Table generated by Excel2LaTeX from sheet 'Table (2)'

		% Table generated by Excel2LaTeX from sheet 'Table (2)'
\begin{tabular}{lcccc|cccc}
			\textbf{Panel A: Forward Distances} \\
      & \textbf{Median } & \textbf{Median } & \textbf{Bootstr. p} & \textbf{Bootstr. p} & \textbf{Mean } & \textbf{Mean } & \textbf{Welch-U p} & \textbf{Welch-U p} \\
\textbf{Announcement} & \boldmath{}\textbf{of $d^+$}\unboldmath{} & \boldmath{}\textbf{of $\tilde{d}^+$}\unboldmath{} & \textbf{left tail} & \textbf{right tail} & \boldmath{}\textbf{of $d^+$}\unboldmath{} & \boldmath{}\textbf{of $\tilde{d}^+$}\unboldmath{} & \textbf{left tail} & \textbf{right tail} \\
\midrule
ADP Employment Change & 20.500 & 34.000 & 0.017* & 1.000 & 37.878 & 55.275 & 5.90E-04*** & 0.999 \\
CPI Core Index SA & 27.250 & 34.000 & 0.198 & 0.967 & 40.315 & 55.282 & 3.74E-03** & 0.996 \\
Change in Nonfarm Payrolls & 27.250 & 34.000 & 0.165 & 0.993 & 49.572 & 55.279 & 0.040* & 0.960 \\
Chicago Purchasing Manager & 33.250 & 40.000 & 0.047* & 0.987 & 53.608 & 60.534 & 0.046* & 0.954 \\
FOMC Rate Decision (Upper Bound) & 22.250 & 35.756 & 4.00E-04*** & 1.000 & 45.150 & 56.674 & 6.18E-04*** & 0.999 \\
Factory Orders & 33.250 & 40.000 & 0.193 & 0.985 & 56.959 & 60.383 & 0.177 & 0.823 \\
Initial Jobless Claims & 34.000 & 34.000 & 0.849 & 0.942 & 53.994 & 55.274 & 0.267 & 0.733 \\
Nonfarm Productivity & 31.917 & 34.000 & 0.253 & 0.758 & 53.381 & 55.269 & 0.400 & 0.600 \\
Underemployment Rate & 13.750 & 34.000 & 0.010* & 0.999 & 28.054 & 55.291 & 8.67E-04*** & 0.999 \\
			\\
			\textbf{Panel B: Backward Distances} \\
			% Table generated by Excel2LaTeX from sheet 'Table (2)'
      & \textbf{Median } & \textbf{Median } & \textbf{Bootstr. p} & \textbf{Bootstr. p} & \textbf{Mean } & \textbf{Mean } & \textbf{Welch-U p} & \textbf{Welch-U p} \\
\textbf{Announcement} & \boldmath{}\textbf{of $d^-$}\unboldmath{} & \boldmath{}\textbf{of $\tilde{d}^-$}\unboldmath{} & \textbf{left tail} & \textbf{right tail} & \boldmath{}\textbf{of $d^-$}\unboldmath{} & \boldmath{}\textbf{of $\tilde{d}^-$}\unboldmath{} & \textbf{left tail} & \textbf{right tail} \\
\midrule
ADP Employment Change & 33.583 & 40.417 & 0.096 & 0.961 & 50.252 & 60.608 & 0.031* & 0.969 \\
CPI Core Index SA & 30.292 & 40.417 & 0.034* & 0.971 & 47.574 & 60.617 & 0.011* & 0.989 \\
Change in Nonfarm Payrolls & 41.917 & 40.417 & 0.783 & 0.225 & 66.198 & 60.683 & 0.903 & 0.097 \\
Chicago Purchasing Manager & 41.208 & 34.083 & 0.961 & 0.039* & 60.121 & 55.374 & 0.892 & 0.108 \\
FOMC Rate Decision (Upper Bound) & 45.083 & 38.409 & 0.864 & 0.136 & 67.152 & 59.092 & 0.972 & 0.028* \\
Factory Orders & 34.250 & 34.167 & 0.670 & 0.412 & 56.896 & 55.493 & 0.530 & 0.470 \\
Initial Jobless Claims & 40.333 & 40.417 & 0.418 & 0.875 & 60.778 & 60.719 & 0.558 & 0.442 \\
Nonfarm Productivity & 40.500 & 40.417 & 0.729 & 0.376 & 63.926 & 60.649 & 0.663 & 0.337 \\
Underemployment Rate & 26.625 & 40.417 & 0.038* & 0.962 & 39.176 & 60.533 & 0.011* & 0.989 \\

\end{tabular}%
	}
\end{table}

\subsection{Nasdaq Nordic markets}

Table \ref{tab:Medians_means_Nordic} shows the left- and right-tailed p-values using bootstrapping and the Welch U-test method Nordic firm-specific scheduled and non-scheduled announcements. Announcement data were collected from Nasdaq OMX Copenhagen, Stockholm, and Helsinki. Both forward and backward waiting times are considered. We use the filtered data set that excludes announcement that had another (scheduled or non-scheduled) announcement in the neighborhood of 6 hours on both sides. 

Table \ref{tab:Medians_means_Nordic}, Panel A, shows that forward distances between scheduled company announcements and return jumps are clearly abnormally low across all the market places. This holds both in terms of medians and means with a great statistical significance. Also the economic significance is high: The mean values of forward distances with actual announcements are approximately half compared to the reference values. The impact of schedule announcement on waiting times is even greater in terms of medians as the median waiting times are around 20 hours in the reference sample and 0.253 hours (15 minutes, which is the length of the jump period) with the actual announcements. 

At the same time, in the case of non-scheduled announcements, forward distances after the announcements are abnormally small with statistical significance at the marketplaces in Copenhagen (Denmark) and Helsinki (Finland), but not at Stockholm (Sweden). Therefore, our results show that releases of unexpected information (non-scheduled announcements) are not reacted to uniformly across Nasdaq Nordic market places, even if they are jointly operated and are based on the same exchange rules.

Regarding backward distances (Panel B), there are no consistent significance of information leakage among three datasets for each Nordic market. In particular, p-values for the alternative hypotheses $\mathbb{E}(d^-) < \mathbb{E}(\tilde{d^-})$ are found significant for scheduled announcements in Copenhagen only. From an economic view-point, this may signal that there is information leakage in the Danish market than the other two Nordic markets, but no clear evidence of information leakage can be provided. 

We found that filtering out confounding events---which is done by excluding announcements that had another (scheduled or non-scheduled) announcement in the neighborhood of 6 hours on both sides---is important. If all the data are used, there is seemingly clear abnormal behavior of the backward differences (results available upon request), but when neighborhood events are filtered out, no clear evidence of information leakage is provided anymore.

\begin{table}[!hp]
	\caption{\textbf{Medians and means of the forward (Panel A) and backward distances (Panel B) for the Nasdaq Nordic data sample.} {\footnotesize \normalfont The results are based on filtered data set that excludes announcements that had another (scheduled or non-scheduled) announcement in the neighborhood of 6 hours on both sides. The Bootstr. p left tail and Bootstr. p right tail are the left- and right-tailed p-values calculated with the bootstrapping method, respectively, and the Welch U-test p left tail and Welch U-test p right tail are the left and right-tailed p-values for the mean values calculated with the Welch U-test (unequal variances t-test for ranked data), respectively.}
		\label{tab:Medians_means_Nordic}}
	\centering
	\resizebox{1\textwidth}{!}{
		\begin{tabular}{lcccc|cccc}
			\multicolumn{7}{l}{\textbf{Panel A: Forward Distances}}\\ \ \\
			& \textbf{Median } & \textbf{Median } & \textbf{Bootstr. p} & \textbf{Bootstr. p} & \textbf{Mean } & \textbf{Mean } & \textbf{Welch U-test p} & \textbf{Welch U-test p} \\
			& \boldmath{}\textbf{of $d^+$}\unboldmath{} & \boldmath{}\textbf{of $\tilde{d}^+$}\unboldmath{} & \textbf{left tail} & \textbf{right tail} & \boldmath{}\textbf{of $d^+$}\unboldmath{} & \boldmath{}\textbf{of $\tilde{d}^+$}\unboldmath{} & \textbf{left tail} & \textbf{right tail} \\
			\hline
			\textbf{Scheduled} &       &       &       &       &       &       &       &  \\
			DAN & 0.253   & 19.530   & 0***  & 1.000 & 12.535   & 29.293   & 5.658E-33*** & 1.000 \\
			SWE & 0.253   & 23.230   & 0***  & 1.000 & 19.410   & 34.866   & 3.347E-30*** & 1.000 \\
			FIN & 0.253   & 26.503   & 0***  & 1.000 & 19.795   & 38.883   & 2.526E-38*** & 1.000 \\
			
			\textbf{Non-Scheduled} &       &       &       &       &       &       &       &  \\
			DAN & 15.499   & 18.928   & 0***  & 1.000 & 25.258   & 28.245   & 2.605E-09*** & 1.000  \\
			SWE & 24.023   & 24.224   & 0.332 & 0.668 & 34.627   & 35.565   & 0.080  & 0.920 \\
			FIN &  25.253   & 26.500   & 0.039* & 0.961 & 38.368   & 39.870   & 2.809E-03** & 0.997  \\
			
			\ \\ \multicolumn{7}{l}{\textbf{Panel B: Backward Distances}}\\ \ \\
			& \textbf{Median } & \textbf{Median } & \textbf{Bootstr. p} & \textbf{Bootstr. p} & \textbf{Mean } & \textbf{Mean } & \textbf{Welch U-test p} & \textbf{Welch U-test p} \\
			& \boldmath{}\textbf{of $d^-$}\unboldmath{} & \boldmath{}\textbf{of $\tilde{d}^-$}\unboldmath{} & \textbf{left tail} & \textbf{right tail} & \boldmath{}\textbf{of $d^-$}\unboldmath{} & \boldmath{}\textbf{of $\tilde{d}^-$}\unboldmath{} & \textbf{left tail} & \textbf{right tail} \\
			\hline
			\textbf{Scheduled} &       &       &       &       &       &       &       &  \\
			DAN & 19.221   & 23.099   & 0.076 & 0.924 & 30.108   & 31.701   & 0.047* & 0.953 \\
			SWE & 25.997   & 25.997   & 0.779 & 0.221 & 36.951   & 38.296   & 0.341 & 0.659 \\
			FIN & 27.997   & 27.958   & 0.528 & 0.472 & 46.852   & 41.916   & 0.867 & 0.133 \\
			
			\textbf{Non-Scheduled} &       &       &       &       &       &       &       &  \\
			DAN &  20.281   & 22.086   & 0.032* & 0.968 & 29.711   & 30.604   & 0.056 & 0.944 \\
			SWE  & 25.997   & 25.997   & 0.944 & 0.056 & 37.748   & 37.664   & 0.533 & 0.467  \\
			FIN &  26.748   & 26.999   & 0.340 & 0.660 & 40.030   & 41.420   & 0.370 & 0.630 \\
			
		\end{tabular}%
	}
\end{table}

\subsection{Specific company announcements in Nordic markets }
This section investigates the statistical association of jumps to five specific important types of company announcements in Danish, Swedish and Finnish markets. Instead of classifying announcements to scheduled or non-scheduled, here they are classified based on the contents. The five selected announcements\footnote{To access the announcements under this classification, visit \url{http://www.nasdaqomxnordic.com/news/companynews}.} are:
\begin{enumerate}
	\item \textbf{Acquisition}: This class of announcements contains news events on the acquisition activity of one company that is interested in another one. Releases related to acquisitions include all the actions and phases belonging to acquisition processes, from intention to closing. For example, one announcement by Nokia on August 7, 2006 in Helsinki was that Nokia did not recommend or endorse a below-market, mini-tender offer of up to 5 million Nokia ADSs from TRC Capital.
	\item \textbf{Change in Board Composition}:
	All the announcements related to personnel changes, resignations, appointments, and retirements in relation to a company’s board or management are included. Proposals and nominees for board/committee members are included as well as constitutive meetings of the board.
	For instance, one announcement by Finnair Oyj on February 28, 2008 was Jaana Tammisto’s appointment as managing director of Finland’s travel bureau.
	
	\item \textbf{Change in Capital Structure}: This family of announcements concerns companies’ decisions regarding changes in capital including changes in capital structure and top managerial levels. Releases are related to share offerings, changes in share capital and votes, subscriptions of shares with options and warrants, and the listing of issued options. All actions and phases are included. The following is an example from Novo Nordisk A/S in Copenhagen on December 27, 2010: Novo Nordisk A/S share repurchase program started.

	\item \textbf{Company Announcement}:
	All the announcements that do not belong to any of the other categories and general announcements concerning a company’s actions are included. Releases in this class include multiple types of information. For example, Elisa’s Annual General Meeting was held on Thursday, March 18, 2010, and the information was released by Nasdaq Helsinki. Furthermore, Nordea Bank AB (publ) released a company announcement on March 09, 2007 from Stockholm announcing the completion of its acquisition of Orgresbank.
	
	\item \textbf{Interim Report}: 
	Interim reports include financial reports from periods shorter than one year. For instance, KONE Corporation’s interim report for January to September 2009 was released on October 20, 2009 at 12:30 p.m. with the following quotation: “KONE further specifies its operating income outlook for 2009. In
	operating income (EBIT), the objective is EUR 580–595 million,
	excluding the one-time cost of EUR 33.6 million, which was booked in
	the second quarter. The previous operating income (EBIT) outlook was
	EUR 570–595 million excluding the one-time cost of EUR 33.6 million.”
	
\end{enumerate}

The above five announcement classes were selected for a number of reason. First, acquisitions are among the most important corporate events in financial markets. The values of the bidder firm and target company normally both change due to the takeover, as does shareholders’ wealth (see \citep{eckbo2008handbook}). Changes in board composition strongly affect managerial performance, which is essential to a firm. The effects of changes in capital structure on asset prices are arguable. Both these news classes---acquisition and change in board composition---are important to financial investors and academia. We also select two comprehensive news classes---company announcements and interim reports. The motivation for selecting these two most frequently released news classes is to investigate the possibility of information leakage and the speed of market reaction in terms of jumps. The other reason is to gain statistical robustness. These five news classes have larger sample sizes than other news.

The main contribution of this analysis to the finance literature is that not only the statistical association of selected news events to detected jumps but also empirical evidence of Nordic market efficiency partially provided in terms of jumps. Since there is vast literature, both theoretical and empirical work, on the mechanism and impacts of merger and acquisition, changes in board composition, and changes in capital structure, we only focus on the market reaction to these particular important announcements in terms of jumps in terms of the forward/backward time distances to/from return jumps. Additionally, two comprehensive families of announcements---company announcements and interim reports---are investigated. To the best of our knowledge, there has been little research directly considering the stock prices reacting to company announcements and interim reports in the NASDAQ Nordic database. %In particular, interim reports have been found in our research to be extremely important information sources due to their strong statistical relationship to jumps in stock prices.

Table \ref{TAB:SelectedNewsMeanTest} reports the Welch U-tests for the equality of means of waiting times (results for medians are available upon request). First, for interim reports empirical forward distances are significantly smaller compared to the corresponding reference ones among all three Nordic markets. This provides evidence on the importance of interim report, which normally is ignored in finance literature. Additionally, results on changes in board also show statistical significance in Danish and Finnish markets, but not in Sweden.\footnote{Notice that for company announcements in Finnish stock markets, the  mean of $d^+$ is lower that of $\tilde{d}$, yet the Welch U-test provides indicates significance in left-tail test. This is possible because Welch U-test considers the {\em ranks} of the observations and not the observations themselves, being insensitive to outliers.} Surprisingly, results for acquisitions are significant with Danish data only. Second, regarding backward distances one interesting observation is that, for all three markets, the waiting times from jumps to the the arrival of company announcements are abnormally distributed with statistical significance. This can indicate information leakage. In particular, the backward waiting times statistically longer compared to the reference ones, which implies that stock prices jump in advance well-before the actual announcements arrive.

\begin{table}[!h]
	\caption{\textbf{Means of waiting times associated to five selected news for Nordic large-cap data.} {\footnotesize \normalfont The results are based on the filtered data set that exclude announcements that had another announcement 
			6 hours on both sides for Danish, Swedish and  Finnish markets. {\#} Obs represents the number of observations. The Welch U-test p left tail and Welch U-test p right tail are the left and right-tailed p-values for the mean values calculated with the Welch U-test (unequal variances t-test for ranked data), respectively.}
		\label{TAB:SelectedNewsMeanTest}}
	\centering
	\resizebox{0.9\textwidth}{!}
	{
		\begin{tabular}{lcccccc}
			
			\textbf{Panel A}& \textbf{\#}\textbf{Obs of } & \textbf{Mean of } & \textbf{\#}\textbf{Obs of} & \textbf{Mean of } & \textbf{Welch U-test } & \textbf{Welch U-test }\\
			\textbf{Forward Distances}& \textbf{ $d$} & \textbf{$d$} & \textbf{ $\tilde{d}$} & \textbf{ $\tilde{d}$} & \textbf{p left tail} & \textbf{p right tail}\\

			\multicolumn{5}{l}{\textbf{ DAN} }\\
			\midrule
		Acquisition &84 &17.85 &7025 &27.22 &2.459E-03** &0.99 \\            
		Changes in board &45 &28.57 &1961 &29.28 &3.989E-02* &0.96 \\           
		Changes in capital &82 &33.94 &6710 &28.78 &0.61 &0.38 \\        
		Company Announcement &759 &24.44 &466731 &26.19 &4.113E-06*** &1.000 \\
		Interim report &157 &11.96 &22508 &28.43 &7.559E-22*** &1.000 \\
			\multicolumn{5}{l}{\textbf{ SWE} }\\
			\midrule
			Acquisition &254 &34.97 &58526 &37.34 &0.13 &0.86 \\             
			Changes in board &176 &34.14 &28044 &33.69 &0.37 &0.62 \\           
			Changes in capital &60 &27.60 &3497 &32.94 &0.24 &0.75 \\        
			Company Announcement &1602 &33.51 &1138323 &31.15 &0.87 &0.12 \\ 
			Interim report &170 &14.24 &26011 &33.54 &3.535E-28*** &1.000 \\ 
			 
			{\textbf{ FIN } }\\ \midrule
			Acquisition &171 &38.62 &27012 &36.62 &0.39 &0.60 \\               
			Changes in board &181 &30.40 &28164 &36.89 &1.436E-03** &0.99 \\           
			Changes in capital &415 &41.91 &127649 &40.20 &0.79 &0.20 \\       
			Company Announcement &1281 &36.16 &855195 &34.60 &7.927E-03** &0.99 \\  
			Interim report &198 &22.73 &35000 &37.50 &7.482E-19*** &1.000 \\    
			\\
			
			\textbf{Panel B}& \textbf{\#}\textbf{Obs of } & \textbf{Mean of } & \textbf{\#}\textbf{Obs of} & \textbf{Mean of } & \textbf{Welch U-test } & \textbf{Welch U-test }\\
			\textbf{Backward Distances}& \textbf{ $d$} & \textbf{$d$} & \textbf{ $\tilde{d}$} & \textbf{ $\tilde{d}$} & \textbf{p left tail} & \textbf{p right tail}\\

			\multicolumn{5}{l}{\textbf{ DAN} }\\
			\midrule
		Acquisition &84 &23.86 &7025 &30.90 &4.349E-03** &0.99 \\           
		Changes in board &45 &22.99 &1961 &29.72 &0.10 &0.89 \\            
		Changes in capital &82 &29.34 &6710 &32.83 &0.26 &0.73 \\       
		Company Announcement &759 &31.08 &466731 &28.59 &0.97 &2.748E-02* \\
		Interim report &157 &29.93 &22508 &30.92 &0.33 &0.66 \\  
		
			\multicolumn{5}{l}{\textbf{ SWE}}\\
			\midrule
			
			Acquisition &254 &36.95 &58526 &39.91 &0.24 &0.75 \\               
			Changes in board &176 &36.71 &28044 &35.59 &0.38 &0.61 \\             
			Changes in capital &60 &40.92 &3497 &36.56 &0.48 &0.51 \\          
			Company Announcement &1602 &36.58 &1138323 &32.31 &1.000 &5.105E-07*** \\
			Interim report &170 &29.44 &26011 &36.46 &6.314E-02 &0.93 \\  
			
			\textbf{ FIN } \\
			\midrule
		Acquisition &171 &36.93 &27012 &38.76 &0.35 &0.64 \\               
		Changes in board &181 &39.67 &28164 &39.91 &0.30 &0.69 \\             
		Changes in capital &415 &39.70 &127649 &41.80 &7.779E-02 &0.92 \\       
		Company Announcement &1281 &40.32 &855195 &35.83 &1.000 &1.203E-05*** \\ 
		Interim report &198 &40.39 &35000 &39.88 &0.67 &0.32 \\    
		\end{tabular} }
\end{table}

\subsection{Relation of sizes of jumps and news events}

This section discusses the statistical characteristics of jump sizes for stock prices associated with scheduled and non-scheduled announcements in the Helsinki, Stockholm, and Copenhagen exchanges. It is important is to have a profound understanding of jump sizes and their relationship to the arrival of announcements as the size of the jump can be considered to measure how important certain unexpected news arrivals are. However, there is barely research about the magnitude of jumps around the news announcements. Modifying the present statistical framework the investigation of for jump sizes instead of waiting times allows us to analyze the relation between the magnitude of jumps and the type of announcements. Here the nearest forward and backward jumps are collected for measuring their sizes. Particularly, the absolute values of these nearest jumps are associated with scheduled and non-scheduled announcements on both forward and backward directions. That is, all jump sizes are the absolute value of detected extremal log returns divided by the corresponding volatility---that is, the $|\mathcal{L}|$, where $\mathcal{L}$ is given by Eq. (\ref{EQ:L}). The reference data is following the same procedure as before (described in Section \ref{subsubsec:data_simulations}) and the reference data  $|\mathcal{\tilde{L}}|$ represents the magnitudes of jumps in the neighborhood of simulated time stamps. Again, Welch U-test is applied, but now to empirical jump sizes and their reference counterparts rather than waiting times.

%The empirical analysis shows that large jumps significantly follow scheduled announcements. In contrast, backward non-scheduled announcements do not contribute to jumps in abnormal sizes. Additionally, negative jumps are found to dominate positive jumps in number and size for both scheduled and non-scheduled announcements.

Table \ref{TAB:JumpSizeMeanTest} presents the mean test results for the sizes of jumps for Nordic markets. We  observe that the mean values of empirical forward jump sizes associated with scheduled announcements are statistically larger than the means of reference samples consistently among the filtered data sets. Similarly, the mean values of empirical jump sizes associated with non-scheduled announcements are statistically larger than the means of the reference data, although this empirical property is not consistent among all data sets as results for Stockholm market place are insignificant. This confirms that in Sweden, indeed, non-scheduled announcements are not reacted significantly both in terms of time distances from announcements to return jumps and sizes of jumps. 

Concerning the jumps that occurred before the announcements, evidence is provided that jumps are abnormally small before scheduled announcements in Denmark and Finland and before non-scheduled announcement in Sweden. Therefore, even if the jumps that occur before the announcements are located normally (see Table \ref{tab:Medians_means_Nordic}), they are abnormally small. However, it is not clear whether this is related to information leakage or not; one could postulate that information leakage would lead to abnormally {\em large} jumps that arrive before the public announcements.

\begin{table}[h]
	\caption{\textbf{Means of normalized jump sizes for Nordic large-cap data. } {\scriptsize  The results are based on all announcements and the filtered data set  that excludes announcements that had another (scheduled or non-scheduled) announcement in the neighborhood of 6 hours on both sides, respectively. {\#} Obs represents the number of observations. The Welch U-test p left tail and Welch U-test p right tail are the left and right-tailed p-values for the mean values calculated with the Welch U-test (unequal variances t-test for ranked data), respectively.}
		\label{TAB:JumpSizeMeanTest}}
	\centering
	\resizebox{0.83\textwidth}{!}
	{
		\begin{tabular}{lcccccc}
			\\
			\multicolumn{7}{l}{\textbf{Panel A: Sizes of jumps that arrived after announcements}}\\
			& \textbf{\#}\textbf{Obs of } & \textbf{Mean of} & \textbf{\#}\textbf{Obs of } & \textbf{Mean of } & \textbf{Welch U-test } & \textbf{Welch U-test}\\
			& \textbf{$|L|$} & \textbf{$|L|$} & \textbf{$\tilde{|L|}$} & \textbf{$\tilde{|L|}$} & \textbf{p left tail} & \textbf{p right tail}\\
			\midrule
            \textbf{Scheduled} &       &       &       &       &       &      \\
			DK &256 &14.44 &65854 &10.89 &1.000 &3.387E-12*** \\  
			SWE &307 &16.37 &94183 &11.59 &1.000 &1.557E-20*** \\   
			FIN &338 &15.52 &114189 &10.41 &1.000 &5.494E-25*** \\    
			\textbf{Non-Scheduled}&       &       &       &       &       &      \\
			DK &1078 &12.78 &1163218 &10.86 &0.99 &4.711E-04*** \\
			SWE &2053 &12.87 &4214741 &11.75 &0.42 &0.57 \\
			FIN &1911 &11.13 &3650026 &10.37 &0.99 &1.603E-03** \\ 
			\\
			\multicolumn{7}{l}{\textbf{Panel B: Sizes of jumps that arrived before announcements}}\\
			\textbf{}& \textbf{\#}\textbf{Obs of } & \textbf{Mean of} & \textbf{\#}\textbf{Obs of } & \textbf{Mean of } & \textbf{Welch U-test } & \textbf{Welch U-test}\\
			\textbf{}& \textbf{$|L|$} & \textbf{$|L|$} & \textbf{$\tilde{|L|}$} & \textbf{$\tilde{|L|}$} & \textbf{p left tail} & \textbf{p right tail}\\
			\midrule
			
			\textbf{Scheduled} &       &       &       &       &       &      \\
			DK &256 &8.958 &65854 &10.53 &1.038E-03** &0.99 \\ 
			SWE &307 &10.37 &94183 &11.29 &0.75 &0.24 \\     
			FIN &338 &10.32 &114189 &10.58 &6.673E-03** &0.99 \\  
			\textbf{Non-Scheduled}&       &       &       &       &       &      \\
			DK &1078 &11.82 &1163218 &10.51 &0.38 &0.61 \\
			SWE &2053 &11.40 &4214741 &11.29 &5.388E-02* &0.94 \\ 
			FIN &1911 &11.25 &3650026 &10.51 &0.75 &0.24 \\     
	\end{tabular} }
\end{table}

\subsection{Discussion}\label{Sec:Discussion}

Our method provided evidence for and against the relation between announcements and jump dynamics, depending on the data set. Some of the p-values are extremely low, for example, for the forward distances with scheduled announcements in the Nordic data, while other p-values are clearly not statistically significant, especially for the backward distances. These high p-values especially for the backward distances with non-scheduled company announcements at Nasdaq Nordic are very important in the validation of our framework; if all the results were statistically significant, one could suspect that a method for generating a reference distribution is not realistic. In this regard, we noticed that it is important to use an empirical distribution, fitted to the empirical announcement time-stamps, to simulate the time-stamps for the reference data sample because the use of uniform distribution in the time-stamp generation with our approach overweighed the proportion of the news during non-trading hours. Moreover, announcements published during trading hours are far from uniformly distributed, and therefore, all the approaches yield skewed results when the empirical distributions are not used (see Figure \ref{FIG:announcement_histograms} for the intraday periodicity/seasonality of announcements). Generally, the use of uniform distributions can lead to p-values that are too low and not realistic. For example, the use of uniform distributions for the announcement time-stamps in the generation of the reference data sets would indicate information leakage with non-scheduled announcements in the Copenhagen exchange data whereas this is not the case with the empirical distributions.

\section{Conclusion}\label{Sec:Conclusion}

In this paper, we provide a non-parametric framework to statistically discover the impact of announcements on the arrival of jumps in stock prices around announcements. The framework can be used to analyze how fast the markets react to the arrival of specific types of announcements in terms of jumps in stock prices and to analyze markets' pre-reactions, possibly induced by information leakage. Also, the magnitude of the jumps can be analyzed. The method was applied to  SPY intraday prices with macro announcements and Nasdaq Nordic equity data with company announcements. 

Some macro announcements, such as the FOMC Rate Decision, are found to contribute to pre- and post-announcement jump dynamics. However, for other announcement releases, such as Factory Orders, we found no evidence of a statistical relation with the jump process of SPY. In addition, scheduled company announcements have clear impacts on the post-announcement jump process across the three Nordic markets. Interestingly, whereas non-scheduled announcements cause jumps in stock prices in the Copenhagen and Helsinki exchange data, no statistical relation is observed with data from the Stockholm exchange. Additionally, we find some evidence for abnormal jump dynamics in equity prices that precedes company announcements on the Copenhagen and Helsinki exchanges, which can indicate information leakage. Overall, this research shows that jumps in stock prices can be driven by past and forthcoming announcements, an empirical observation that could be used to improve existing jump models in risk management and option pricing.

Finally, it would be interesting in future research to include information about the signs and sizes of jumps to our framework to answer if the positive and negative revises of investors' (biased) beliefs\footnote{Recently, \cite{engelberg2016anomalies} find that anomaly returns are substantially higher on earnings announcement and corporate news days.} happen in a non-gradual way, which could be observed from a joint distribution of the waiting-times and the sizes of jumps around news events.

\section*{References}
\bibliography{references}

\end{document}